\documentclass[reprint,amsmath,amssymb,aps,nofootinbib,amsart]{revtex4-2}
\pdfoutput=1
\usepackage{graphicx,amsmath,amssymb,amsthm}
\usepackage{hyperref}
\usepackage{xcolor}
\usepackage{ulem}
\usepackage{thmtools, thm-restate}

\begin{document}
\def\nn{\nonumber}
\def\kc#1{\left(#1\right)}
\def\kd#1{\left[#1\right]}
\def\ke#1{\left\{#1\right\}}
\newcommand\beq{\begin{equation}}
\newcommand\eeq{\end{equation}}
\renewcommand{\Re}{\mathop{\mathrm{Re}}}
\renewcommand{\Im}{\mathop{\mathrm{Im}}}
\renewcommand{\b}[1]{\mathbf{#1}}
\renewcommand{\c}[1]{\mathcal{#1}}
\renewcommand{\u}{\uparrow}
\renewcommand{\d}{\downarrow}
\newcommand{\be}{\begin{equation}}
\newcommand{\ee}{\end{equation}}
\newcommand{\bsigma}{\boldsymbol{\sigma}}
\newcommand{\blambda}{\boldsymbol{\lambda}}
\newcommand{\Tr}{\mathop{\mathrm{Tr}}}
\newcommand{\sgn}{\mathop{\mathrm{sgn}}}
\newcommand{\sech}{\mathop{\mathrm{sech}}}
\newcommand{\diag}{\mathop{\mathrm{diag}}}
\newcommand{\Pf}{\mathop{\mathrm{Pf}}}
\newcommand{\half}{{\textstyle\frac{1}{2}}}
\newcommand{\sh}{{\textstyle{\frac{1}{2}}}}
\newcommand{\ish}{{\textstyle{\frac{i}{2}}}}
\newcommand{\thf}{{\textstyle{\frac{3}{2}}}}
\newcommand{\SUN}{SU(\mathcal{N})}
\newcommand{\N}{\mathcal{N}}

\newcommand{\haoyu}{\textcolor{red}}
\newcommand{\hao}{\textcolor{blue}}
\newcommand{\yasu}{\textcolor{orange}}

\title{An Information Paradox and Its Resolution in de~Sitter Holography}
\author{Hao Geng}
\affiliation{Department of Physics, University of Washington, Seattle, WA 98195, USA}
\author{Yasunori Nomura}
\affiliation{Berkeley Center for Theoretical Physics, Department of Physics, 
University of California, Berkeley, CA 94720, USA; \\
Theoretical Physics Group, Lawrence Berkeley National Laboratory, Berkeley, CA 94720, USA; \\
Kavli Institute for the Physics and Mathematics of the Universe 
 (WPI), UTIAS, The University of Tokyo, Kashiwa, Chiba 277-8583, Japan}
\author{Hao-Yu Sun}
\affiliation{Theory Group, Department of Physics, University of Texas, Austin, TX 78712, USA.}
\preprint{\today}
\begin{abstract}
We formulate a version of the information paradox in de~Sitter spacetime and show that it is solved by the emergence of entanglement islands in the context of the DS/dS correspondence; in particular, the entanglement entropy of a subregion obeys a time-dependent Page curve. Our construction works in general spacetime dimensions and keeps the graviton massless. We interpret the resulting behavior of the entanglement entropy using double holography. It suggests that the spatial distribution of microscopic degrees of freedom depends on descriptions, as in the case of a black hole. In the static (distant) description of de~Sitter (black hole) spacetime, these degrees of freedom represent microstates associated with the Gibbons-Hawking (Bekenstein-Hawking) entropy and are localized toward the horizon. On the other hand, in a global (effective two-sided) description, which is obtained by the quantum analog of analytic continuation and is intrinsically semiclassical, they are distributed uniformly and in a unique semiclassical de~Sitter (black hole) vacuum state.
\end{abstract}

\maketitle

\section{Introduction}

Recent development in AdS/CFT holography provides important new elements to our understanding of the geometrization of the entanglement structure of the boundary conformal field theory, called \textit{entanglement islands}~\cite{Penington:2019npb,Almheiri:2019psf,Almheiri:2019hni,Almheiri:2019yqk,Rozali:2019day,Chen:2019uhq,Almheiri:2019psy,Penington:2019kki,Almheiri:2019qdq,Geng:2020qvw,Gautason:2020tmk,Hashimoto:2020cas,Hartman:2020swn,Hollowood:2020cou,Krishnan:2020oun,Chen:2020uac,Krishnan:2020fer,Chen:2020tes,Hartman:2020khs,Balasubramanian:2020coy,Balasubramanian:2020xqf,Sybesma:2020fxg,Ling:2020laa,Matsuo:2020ypv,Goto:2020wnk,Caceres:2020jcn,Deng:2020ent,Karananas:2020fwx,Wang:2021woy,Verheijden:2021yrb,Kawabata:2021hac}.
They play an important role in proposed resolutions of various information paradoxes, including Hawking's original one for an evaporating black hole~\cite{Hawking:1976ra}. (For recent reviews, see Refs.~\cite{Almheiri:2020cfm,Liu:2020rrn,Nomura:2020ewg,Raju:2020smc}.) A general lesson is that we can compute the Page curve of black hole radiation~\cite{Page:1993wv} by coupling the black hole to a boundary bath system which absorbs radiation. The Page curves calculated in various contexts are all consistent with unitarity, a fundamental postulate of quantum mechanics.

The central technique behind these calculations is the so-called \textit{quantum extremal surface} (QES) prescription~\cite{Engelhardt:2014gca,Faulkner:2013ana}. This prescription is fully quantum and requires calculation of entanglement entropies in bulk curved spacetime, which is a hard task in general. However, if the quantum field theory in the bulk is holographic itself, then this quantum prescription can be realized as a classical prescription in a higher dimensional spacetime~\cite{Almheiri:2019psy}, in which we only need to calculate a classical extremal area surface~\cite{Ryu:2006bv,Hubeny:2007xt,Lewkowycz:2013nqa} under specific boundary conditions near the boundary of the higher dimensional spacetime~\cite{Geng:2020fxl}. Models having this property are dubbed as \textit{doubly holographic}, and they can be naturally constructed in Karch-Randall braneworld~\cite{Karch:2000ct,Karch:2000gx,Takayanagi:2011zk,Karch:2020iit} which uses subcritical branes in the original Randall-Sundrum framework~\cite{Randall:1999vf}.

While it is possible to construct models in more than $2+1$ spacetime dimensions in which a time-dependent Page curve is obtained by coupling a bulk black hole spacetime to a non-gravitational boundary bath~\cite{Penington:2019npb}, such models necessarily lead to a massive graviton~\cite{Geng:2020qvw}.
This is most easily seen in the context of the Karch-Randall braneworld, in which a $d$-dimensional AdS spacetime is modelled as a subcritical brane in a $(d+1)$-dimensional AdS spacetime and the bath is the conformal boundary of this AdS$_{d+1}$. From the higher dimensional ($(d+1)$-dimensional) point of view, the subcritical brane cuts off the portion of the spacetime behind it, but the remaining part is still non-compact. This makes the would-be massless graviton modes localized on the subcritical brane non-normalizable, leaving only massive graviton modes to be physical.

In an attempt to rectify this issue, Ref.~\cite{Geng:2020fxl} considered a setup in which another subcritical brane cuts off the leftover conformal boundary. This makes the extra dimension compact and massless graviton modes survive. In this case, however, the bath is a portion of the second subcritical brane, which is gravitating by itself. This motivated Ref.~\cite{Geng:2020fxl} to apply the same prescription to both the first and second branes, resulting in a time-independent ``Page curve''.

In this paper, we take the view that to obtain the standard Page curve, we must focus on fine-grained entropy of semiclassical Hawking radiation in a weakly gravitating regime, and not the fine-grained entropy of the full microscopic degrees of freedom. We postulate that this can be done by fixing the boundary of the region on the second brane for which the entanglement entropy is calculated, rather than determining it by extremization.%
\footnote{This has also been suggested by Pratik Rath. For an earlier discussion on this issue and the graviton mass, see Refs.~\cite{Krishnan:2020oun,Krishnan:2020fer}.}
For a black hole on an AdS brane, this indeed reproduces features of the time-dependent Page curve. We suspect that with this prescription, the non-factorizable nature of the quantum gravitational Hilbert space identified in Ref.~\cite{Raju:2020smc} does not play a major role, although more detailed studies are needed to obtain a definite conclusion.

To further test the feasibility of the picture described above, in this paper we perform an analogous calculation in de~Sitter (dS) spacetime. We first formulate a version of the information paradox about entanglement entropy in dS spacetime, in analogy with that proposed for AdS black holes in Ref.~\cite{Hartman:2013qma}. The entanglement entropy and fast scrambling dynamics of dS quantum gravity have been studied in various contexts~\cite{Nomura:2011dt,Maldacena:2012xp,Nomura:2017fyh,Aalsma:2020aib,Geng:2020kxh,Haque:2020pmp}. In this paper, we work in the context of the \textit{DS/dS correspondence}~\cite{Karch:2003em,Alishahiha:2004md}, which is a mathematically well-defined holographic framework for dS quantum gravity.

We embed the paradox into the DS/dS context and see if it is resolved by entanglement islands. There are works on entanglement islands in dS spacetime using lower dimensional models of quantum gravity~\cite{Chen:2020tes,Hartman:2020khs,Balasubramanian:2020xqf,Sybesma:2020fxg}. In this paper, we analyze the problem in general dimensions. Using double holography in the DS/dS framework, we find that the standard, time-dependent Page curve is obtained while preserving the masslessness of the graviton. In addition to increasing the feasibility of the postulated island prescription in a weakly gravitating regime, we find it interesting itself that a time-dependent Page curve is obtained in dS spacetime. We will also find that the resulting picture is consonant with the interpretation of maximally extended spacetime in quantum gravity developed in Refs.~\cite{Nomura:2018kia,Nomura:2019qps,Nomura:2019dlz,Nomura:2020ska}.

This paper is organized as follows. In Section~\ref{sec:DS-dS}, we briefly review the DS/dS correspondence. We then formulate a version of the information paradox in dS spacetime in Section~\ref{sec:paradox}, which we embed into the context of the DS/dS correspondence in Section~\ref{sec:embedding}. In Section~\ref{sec:resol}, we will see how the paradox is resolved because of the emergence of entanglement islands. In Section~\ref{sec:interp}, we discuss deeper implications of our result for the structure of spacetime in quantum gravity. Finally, conclusions are given in Section~\ref{sec:concl}.

\section{The DS/\texorpdfstring{\MakeLowercase{D}}{d}S Correspondence}
\label{sec:DS-dS}

The DS/dS correspondence is most easily understood as a deformation of the AdS/CFT correspondence~\cite{Maldacena:1997re,Gubser:1998bc,Witten:1998qj} as follows. Both a $(d+1)$-dimensional de~Sitter space (dS$_{d+1}$) and a $(d+1)$-dimensional anti-de~Sitter space (AdS$_{d+1}$) can be radially foliated by $d$-dimensional dS slices, but with slightly different warp factor:
\begin{equation}
    ds^{2}_{\text{(A)dS}_{d+1}}=dr^2+\sin(\text{h})^2\!\left(\frac{r}{\ell}\right) ds^{2}_{\text{dS}_{d}},
\label{eq:metric}
\end{equation}
where $\ell$ is the curvature length and we call $r$ the radial coordinate (for AdS$_{d+1}$, $r\in(-\infty,\infty)$, and for dS$_{d+1}$, $r\in[0,\pi \ell]$).
For dS spacetime, the spacetime region covered by these coordinates is called a \textit{DS/dS patch}.

The ultraviolet (UV) conformal boundary of AdS$_{d+1}$ is living at $r=\pm\infty$ and the infrared (IR) bulk point is near $r=0$. We can see that when $r\to0,\pi\ell$, the metric of dS$_{d+1}$ behaves in the same way as that of AdS$_{d+1}$ at $r\to0$. The standard lessons of holography then tell us that the holographic dual field theories of quantum gravity in dS$_{d+1}$ and AdS$_{d+1}$ have the same IR fixed point. However, their UV behaviors are very different. For AdS$_{d+1}$, it is UV completed to a conformal field theory (CFT) living on its conformal boundary. But for dS$_{d+1}$, it is not UV-complete and its field theory dual consists of two UV-cutoff CFTs, coupled to each other via dynamical gravity living on the central slice $r=\frac{\pi}{2}\ell$~\cite{Karch:2003em}. This central slice can be thought of as a built-in Randall-Sundrum brane and the graviton localized on it is massless~\cite{Karch:2003em}.

Summarizing, quantum gravity in dS$_{d+1}$ is dual to two cutoff CFTs living on dS$_{d}$ coupled to each other via dynamical gravity~\cite{Alishahiha:2004md}. In this context, questions regarding entanglement entropy, how the duality works, and its implications for the underlying fast scrambling dynamics have been studied in recent years, where several important progresses have been made \cite{Dong:2018cuv,Gorbenko:2018oov,Geng:2019bnn,Lewkowycz:2019xse,Geng:2019zsx,Geng:2019ruz,Geng:2020kxh}.

\section{The Paradox}
\label{sec:paradox}

We now present an apparent paradox involving dS spacetime, which we will resolve later using DS/dS double holography.

\begin{figure}
\centering
  \includegraphics[width=8.5cm]{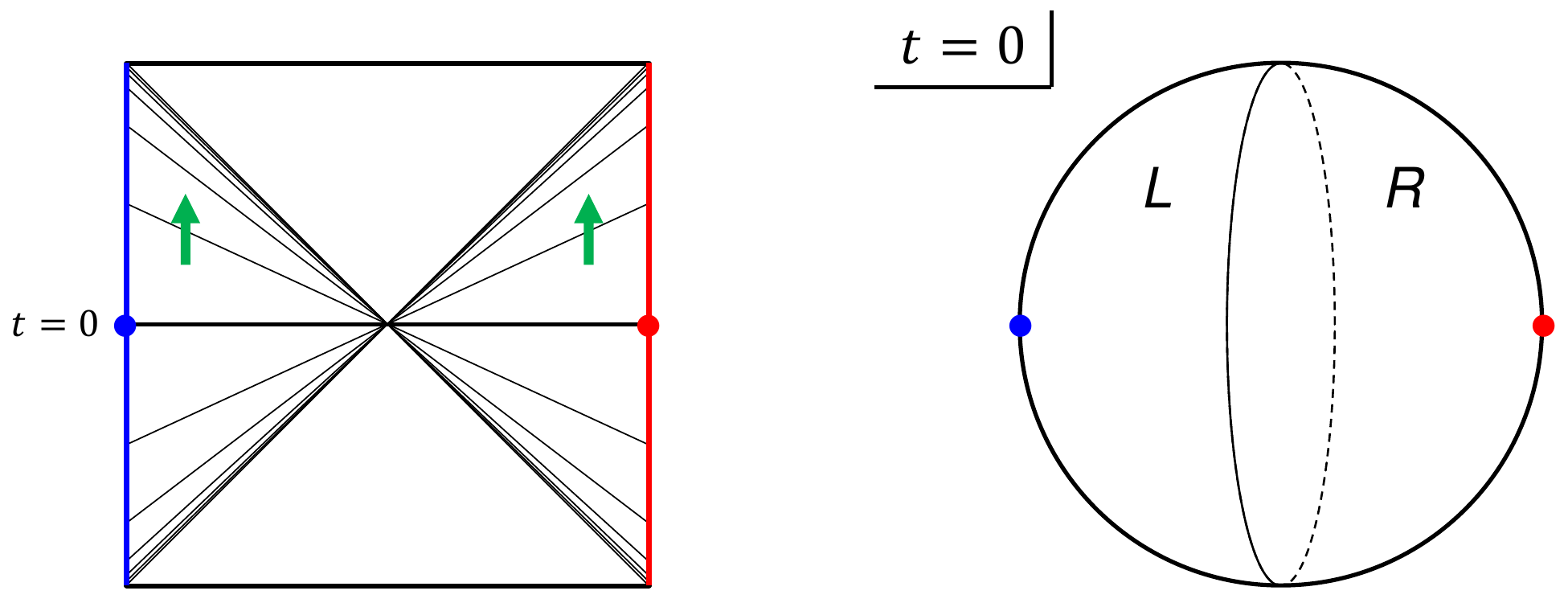}
\caption{\small{\textit{The Penrose diagram of dS spacetime in the extended static patch (left) and its spatial section at $t=0$ (right). The blue and red dots represent the north and south poles at $t=0$. The green arrows represent the direction of time evolution.}}}
\label{fig:ext-st}
\end{figure}

Consider two dS$_{d}$'s. Each of these is dS$_{d}$ in its extended static patch, which is the union of two standard static patches associated with the north and south poles (Fig.~\ref{fig:ext-st}).
We adopt the metric~\cite{Geng:2020kxh}
\begin{equation}
    ds^2=-\cos^2\beta\text{ } dt^2+d\beta^2+\sin^{2}\beta\text{ } d\Omega_{d-2}^2.\label{eq:metric1}
\end{equation}
Here, we have set the curvature length to be one, and $\beta\in[0,\pi]$ for $d \geq 3$ and $\beta \in [0,2\pi)$ for $d=2$. (The difference of the range of $\beta$ between $d=2$ and $d\geq3$ comes from the fact that we have taken the angular direction consisting of a point, not two points, in $d=2$.)

Let us call the two dS spacetimes dS$_d^1$ and dS$_d^2$. We let them interact through some intersecting defects. We impose transparent boundary conditions there, so that radiation can flow from one dS$_d$ to the other. We assume that dS$_d^1$ is gravitating and potentially have some matter fields, and that dS$_d^2$ supports a fast scrambling field theory system.%
\footnote{For the purpose of formulating the paradox, we need not consider that dS$_d^2$ is gravitating. Embedding it into DS/dS, however, requires dS$_d^2$ to be gravitating as dS$_d^1$.}
We take the state of the system in dS$_d^2$ at $t=0$ to be in the thermofield double state between the two hemispheres denoted by $L$ and $R$ in Fig.~\ref{fig:ext-st}.

On the Penrose diagram of a dS$_{d}$ in Fig.~\ref{fig:ext-st}, the parts $L$ and $R$ live on different sides of the horizon, so that the timelike Killing vector $\frac{\partial}{\partial t}$ in Eq.~(\ref{eq:metric1}) goes opposite on the two sides. We break this time translational symmetry, for both dS$_{d}^{1}$ and dS$_{d}^{2}$, by letting the time to go up for both sides as indicated by the green arrows in Fig.~\ref{fig:ext-st}.
With this, the dynamics of the fast scrambling system on dS$_d^2$ is nontrivial.

\begin{figure}
\centering
  \includegraphics[width=8.5cm]{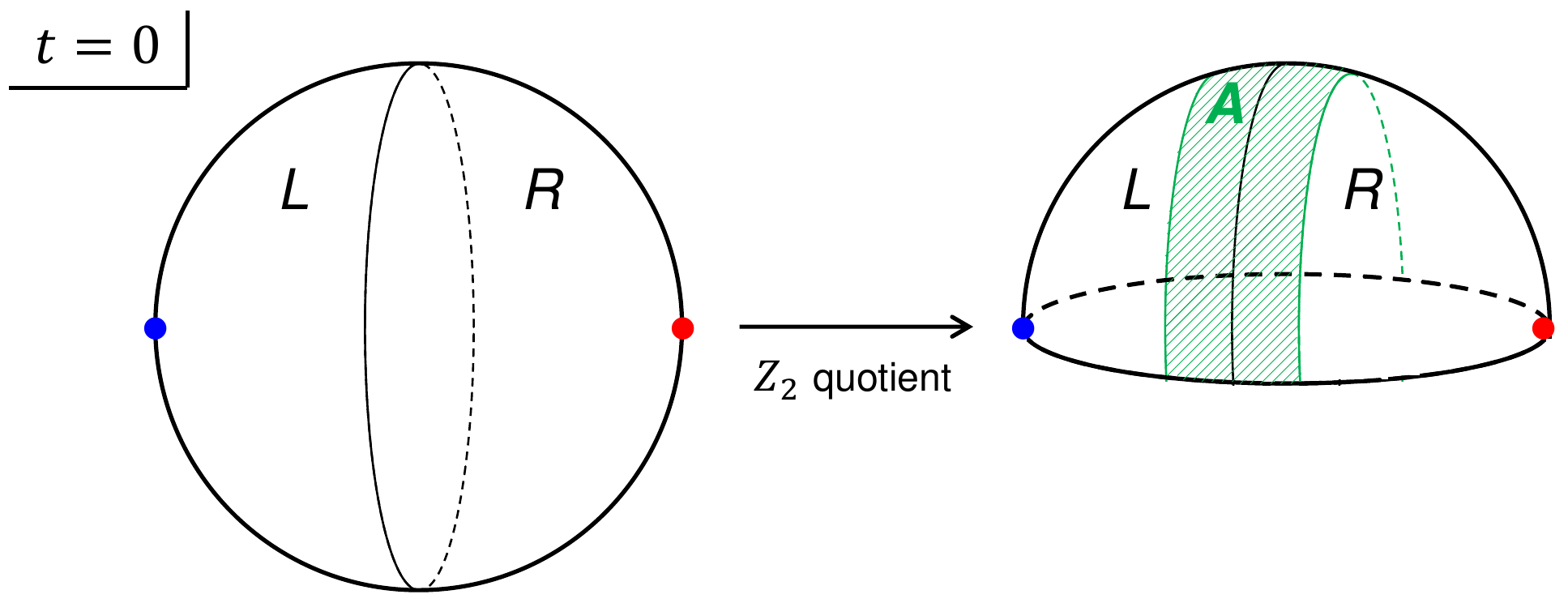}
\caption{\small{\textit{The $\mathbb{Z}_2$ orbifolding removes a half of dS$_d^2$. The region $A$, of which we calculate the entanglement entropy, at $t=0$ is depicted as the green shaded region on the right panel.}}}
\label{fig:Z2}
\end{figure}

For the purpose of embedding the setup into DS/dS later, we take a $\mathbb{Z}_{2}$ quotient of dS$_{d}^{2}$ with respect to a surface that goes through both the north and south poles (and thus is orthogonal to the horizon separating $L$ and $R$ regions); see Fig.~\ref{fig:Z2}. The region of which we calculate the entanglement (von~Neumann) entropy, $A$, is an interior region of this $\mathbb{Z}_2$ quotient, which we take to be symmetric with respect to the horizon, as depicted by the green shaded region in Fig.~\ref{fig:Z2}. The $\mathbb{Z}_2$ quotient dS$_d^2$ space described here plays the role of the bath in Refs.~\cite{Penington:2019npb,Almheiri:2019psf,Almheiri:2019hni,Almheiri:2019yqk,Rozali:2019day,Chen:2019uhq,Almheiri:2019psy,Penington:2019kki,Almheiri:2019qdq}.

We assume that the two spaces dS$_d^1$ and dS$_d^2/\mathbb{Z}_2$ have the same temperature, but they are not significantly entangled with each other at $t=0$.
We then time evolve both dS$_d^1$ and dS$_d^2/\mathbb{Z}_2$ as described above; see the green arrows in Fig.~\ref{fig:ext-st}.
This makes the volume of the region $A$, specified by the coordinate $\beta$, increase in time; it also increases entanglement between dS$_d^1$ and dS$_d^2/\mathbb{Z}_2$ as well as that between $A$ and its complement $\bar{A}$ on dS$_d^2/\mathbb{Z}_2$.
Given that the dynamics of the system on dS$_d^2/\mathbb{Z}_2$ is fast scrambling, the semiclassical expectation is that the entanglement entropy of $A$, which we denote by $S_A$, grows indefinitely in time.

In quantum gravity, however, the Hilbert space dimension of dS spacetime is expected to be finite~\cite{Banks:2000fe,Witten:2001kn}, so such an indefinite increase of entropy would lead to an information paradox.
The correct time dependence of $S_A$ must be the Page curve, i.e., its value must be saturated at late times by the finite Hilbert space dimension of the union of dS$_d^1$ and $\bar{A}$ (or of $A$).

We note that it is important for dS$_{d}^{2}$ to be in the extended static patch of Eq.~(\ref{eq:metric1}), since it is what leads to the increase of the volume of $A$.
The same, however, is not true for dS$^{1}_{d}$; here we put it in this patch simply for convenience of demonstrating the paradox.
Indeed, as we will see in the next section, utilizing double holography requires us to put dS$^{1}_{d}$ in the {\it DS/dS extended static patch}, rather than the extended static patch.
The time evolution of dS$^{1}_{d}$ will then be that of the DS/dS patch.

\section{Embedding the Paradox into DS/\texorpdfstring{\MakeLowercase{D}}{d}S}
\label{sec:embedding}

The first step to embed the setup in the previous section into the DS/dS context is to identify the fast scrambling system on dS$_{d}^{2}$ as the dual of quantum gravity in dS$_{d+1}$. This implies that dS$_{d}^{2}$ is on the central slice of this dS$_{d+1}$, i.e., $r = \frac{\pi}{2}\ell$ in Eq.~(\ref{eq:metric}), where $ds^{2}_{\text{dS}_{d}}$ should now be described by the extended static patch metric in Eq.~(\ref{eq:metric1}). The full metric is now in the DS/dS$_{\text{extended static}}$ patch considered in Ref.~\cite{Geng:2020kxh}, given by
\begin{equation}
  ds^{2}_{\text{dS}_{d+1}} = dr^2 + \sin^2\!r \left( -\cos^2\!\beta\, dt^2 + d\beta^2 + \sin^2\!\beta\, d\Omega_{d-2}^2 \right).
\label{eq:metric2}
\end{equation}
Here and below, we set $\ell = 1$ for convenience.

We model dS$_{d}^{1}$ as a Randall-Sundrum brane living on a slice ``orthogonal'' to that on which dS$_{d}^{2}$ lives. We call this slice $Q$. For $d \geq 4$,
\begin{equation}
  d\Omega_{d-2}^2 = d\chi^2+\sin^2\!\chi\, d\Omega_{d-3}^{2} \quad (0 \leq \chi \leq \pi),
\end{equation}
and $Q$ is the $\chi= \frac{\pi}{2}$ slice. For $d = 3$,
\begin{equation}
  d\Omega_{1}^2 = d\chi^2 \quad (0 \leq \chi < 2\pi),
\end{equation}
and $Q$ is at $\chi = 0$ and $\pi$; note that $Q$ as well as its intersection with dS$_{d}^{2}$ are still single connected surfaces---the $\chi = 0$ and $\pi$ components touch at $\beta = 0$ and $\pi$. Finally, for $d = 2$, $Q$ lives at $\beta = 0$ and $\pi$; in this case, $Q$ is a single connected slice with the $\beta = 0$ and $\pi$ components touching at $r = 0$ and $\pi$, but the intersection of $Q$ and dS$_{d}^{2}$ consists of two disconnected points.

It is easy to see that the geometry of $Q$ is dS$_{d}$ in the DS/dS$_{\text{extended static}}$ patch. This tells us that the bulk gravitational action is given by the Einstein-Hilbert action and boundary terms associated with $Q$:
\begin{equation}
\begin{split}
    S=\frac{1}{16\pi G_{d+1}} & \int\!d^{d+1}x\,\sqrt{-g}(R-2\Lambda)\\
    &-\frac{1}{8\pi G_{d+1}} \int_{Q}\!d^{d}x\,\sqrt{-h}(K-T),
    \end{split}
\end{equation}
where $\Lambda$ is the cosmological constant, $G_{d+1}$ is Newton's constant; $h_{ab}$ is the induced metric on the $Q$ brane, $K$ is the trace of its extrinsic curvature, and $T$ is the brane tension. The boundary condition for the metric fluctuation near $Q$ is of Neumann type, and hence the equation of motion splits into bulk and boundary parts. The bulk part is the vacuum Einstein equation with a cosmological constant, and it is satisfied by the dS$_{d+1}$ in Eq.~(\ref{eq:metric2}) for $\Lambda = \frac{d(d-1)}{2}$. The boundary equation of motion determines the brane tension by~\cite{Kraus:1999it,Fujita:2011fp}
\begin{equation}
    K_{ab}=(K-T)h_{ab}.
\end{equation}
Since it is easily computed that $K_{ab}=0$ for $Q$, the brane $Q$ is tensionless and not fluctuating~\cite{Randall:1999ee,Karch:2020flx}.

The fact that the slice $Q$ is a tensionless brane and non-fluctuating implies that the bulk dS$_{d+1}$ is actually a $\mathbb{Z}_{2}$ quotient with $Q$ as its boundary, where we impose Neumann boundary conditions for bulk fields. This $\mathbb{Z}_{2}$ orbifolding is the $\mathbb{Z}_{2}$ identification of dS$_{d}^{2}$ discussed in Section~\ref{sec:paradox}, i.e.,\  $\mbox{dS}_{d}^{2} \rightarrow \mbox{dS}_{d}^{2}/\mathbb{Z}_{2}$ depicted in Fig.~\ref{fig:Z2}. The two dS systems dS$_{d}^{1}$ and dS$_{d}^{2}$ intersect at the fixed point of this $\mathbb{Z}_{2}$ orbifolding of dS$_{d}^{2}$.

We see that we have successfully embedded the paradox formulated in the previous section into DS/dS. The entanglement entropy of a region $A$ ($\subset \mbox{dS}_{d}^{2}/\mathbb{Z}_{2}$) is expected to increase in time (beyond the Page transition point), which can now be seen as the Hartman-Maldacena surface with increasing area~\cite{Hartman:2013qma} in the dS$_{d+1}$ bulk. In order to terminate this increase, some other element must come in. We stress that the graviton in the current setup is massless, since the holographically generated space is compact.

We finally note that this setup can be viewed as realizing double holography, since the dS$_{d}^{1}$ system on $Q$ itself can be dualized to a dS$_{d-1}$ using the DS/dS correspondence. This dual system then lives on the the boundary of the $\mbox{dS}_{d}^{2}/\mathbb{Z}_{2}$ system (the horizontal great circle on the right panel of Fig.~\ref{fig:Z2}). From this viewpoint, the picture of dS$_{d+1}$ is obtained by double holography, analogous to the setup for AdS black holes in Ref.~\cite{Geng:2020qvw}.

\section{Resolution of the Paradox and a Page Curve}
\label{sec:resol}

To see the picture described above more explicitly, here we focus on the $d = 2$ case. We will see how the paradox manifests itself as a growing bulk entangling surface as well as how it is resolved. The same conclusions apply to the higher dimensional case as well.

\begin{figure}
\centering
  \includegraphics[width=4.5cm]{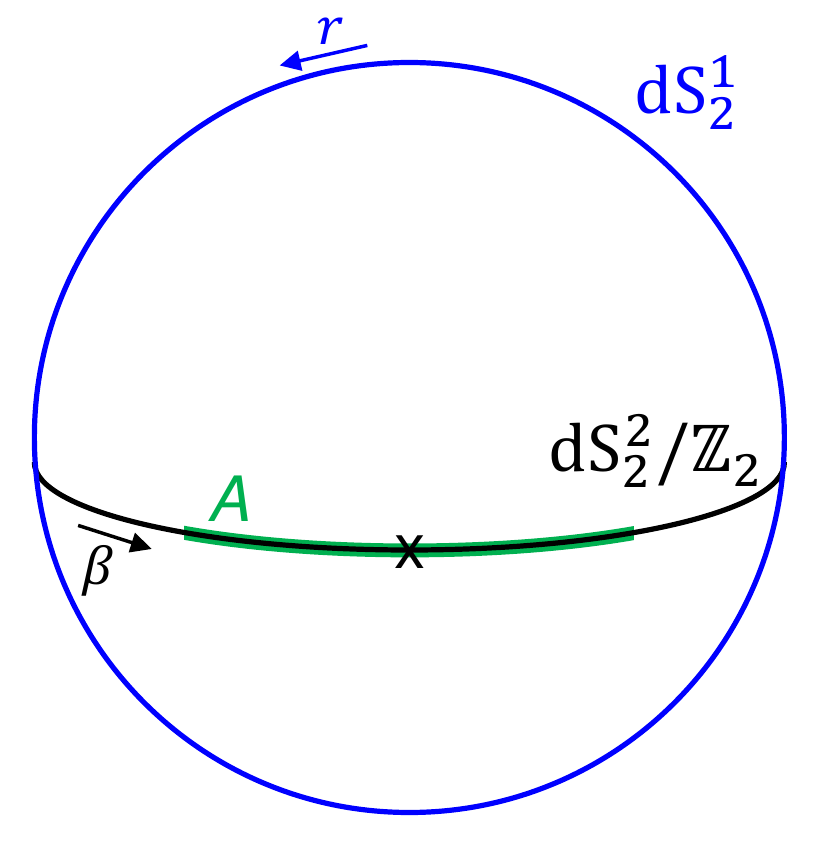}
\caption{\small{\textit{The spatial section at $t = 0$ for $d=2$. The two intersecting dS systems are depicted as the blue circle (dS$_2^1$) and black semicircle ($\mbox{dS}_{2}^{2}/\mathbb{Z}_{2}$). The green segment represents region $A$, of which we calculate the entanglement entropy, and the cross represents the horizon at $\beta = \frac{\pi}{2}$. The holographic bulk space ($\mbox{dS}_{3}/\mathbb{Z}_{2}$) is the hemisphere bounded by dS$_2^1$.}}}
\label{fig:setup}
\end{figure}

For $d=2$, the coordinate $\beta$ goes from 0 to $\pi$ (after the $\mathbb{Z}_{2}$ quotient), and the horizon is located at $\beta = \frac{\pi}{2}$. The field theory system of $\mbox{dS}_{2}^{2}/\mathbb{Z}_{2}$ lives on the $r = \frac{\pi}{2}$ slice, with $\beta = \frac{\pi}{2}$ being the bifurcation surface. The spatial section of the geometry at $t = 0$ is depicted in Fig.~\ref{fig:setup}. The two dS systems dS$_{2}^{1}$ and $\mbox{dS}_{2}^{2}/\mathbb{Z}_{2}$ are a circle and a semicircle, respectively, and the bulk dS$_{3}$ is a hemisphere bounded by dS$_{2}^{1}$.

What about time evolution? As discussed in Section~\ref{sec:paradox}, we evolve the left ($\beta \in [0,\frac{\pi}{2}]$) and right ($\beta \in [\frac{\pi}{2},\pi]$) portions of $\mbox{dS}_{2}^{2}/\mathbb{Z}_{2}$ in the same and opposite directions of the timelike Killing vector, respectively (or vice versa). We consider the time evolution of the entanglement entropy $S_A$ of region $A$, which we take to be $\beta \in [\beta_*,\pi-\beta_*]$ ($0 < \beta_* < \frac{\pi}{2}$) on the $r = \frac{\pi}{2}$ slice. The region $A$ is depicted by the green line segment in Fig.~\ref{fig:setup}.

For this purpose, let us consider the geodesic in dS$_3$ between the two end points of $A$ as a function of time. In terms of the embedding space coordinate for the DS/dS$_{\text{extended static}}$ patch, these two points are located at
\begin{equation}
    \begin{split}
        X_{0}&=\cos\beta_* \sinh t \\
        X_{1}&=0\\
        X_{2}&=\cos\beta_* \cosh t \\
        X_{3}&=\sin\beta_*
    \end{split}
\end{equation}
and
\begin{equation}
    \begin{split}
        X_{0}'&=\cos(\pi-\beta_*)\sinh(-t)=\cos\beta_*\sinh t\\
        X_{1}'&=0\\
        X_{2}'&=\cos(\pi-\beta_*)\cosh(-t)=-\cos\beta_*\cosh t\\
        X_{3}'&=\sin(\pi-\beta_*)=\sin\beta_*,
    \end{split}
\end{equation}
respectively (see Ref.~\cite{Geng:2020kxh}). Then the geodesic distance connecting these points, which would give the entanglement entropy of $A$ through the holographic formula, is given by
\begin{equation}
\begin{split}
  D &= \arccos(-X_{0}X_{0}'+X_{1}X_{1}'+X_{2}X_{2}'+X_{3}X_{3}')\\
    &= \arccos(1-2\cos^{2}\!\beta_*\cosh^{2}\!t).
    \end{split}
\label{eq:growth}
\end{equation}
This monotonic growth in time is consistent with the fact that dS spacetime is a fast scrambling system.

However, the entanglement entropy of $A$ given by Eq.~(\ref{eq:growth}) through the holographic formula, $S_A = \frac{D}{4G_3}$, is problematic, since it becomes imaginary at $t = t_c \equiv \big|{\rm arccosh}\frac{1}{\cos\beta_*}\big|$. In fact, for $t > t_c$ the expression in Eq.~(\ref{eq:growth}) is inapplicable, and the geodesic between the two end points of $A$ stops existing.

\begin{figure}
\centering
\includegraphics[width=8.1cm]{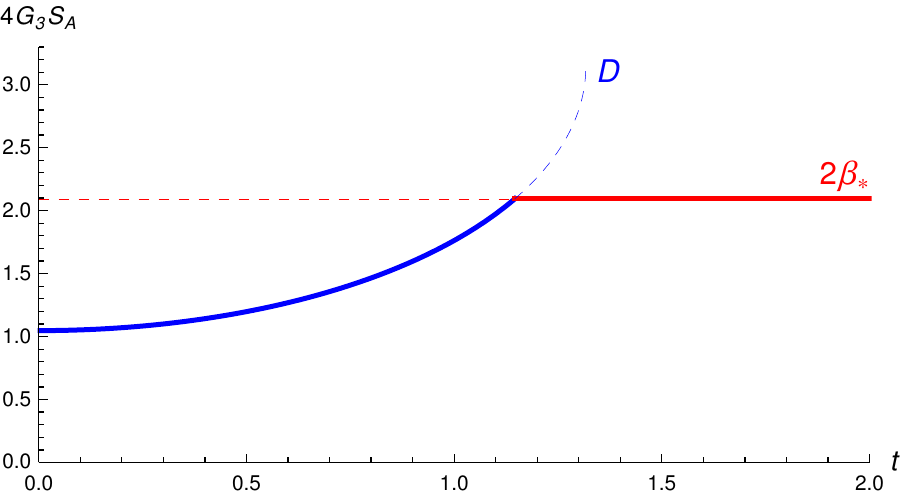}
\caption{\small{\textit{We plot $4 G_3 S_A$ for $\beta_* = \frac{\pi}{3}$ as an example. The blue and red curves represent $D$ and $2\beta_*$, respectively; see Eq.~(\ref{eq:S_A}). The actual $4 G_3 S_A$ is given by the solid part of the curves.}}}
\label{fig:resolution}
\end{figure}

The existence of another QES, however, solves this issue. Since the dS$_2^1$ brane (on the $\beta = \{0,\pi\}$ circle) can be viewed as arising holographically from two dS$_1$'s on $(r,\beta) = (\frac{\pi}{2},0)$ and $(\frac{\pi}{2},\pi)$, a QES for calculating $S_A$ can end on it~\cite{Almheiri:2019hni,Geng:2020fxl}. After the extremization, we find that the relevant QES is the complement of $A$ on $r = \frac{\pi}{2}$, i.e.,\ $\bar{A}$ on the equal-time hypersurface of $\mbox{dS}_{2}^{2}/\mathbb{Z}_{2}$, whose area (length) is simply $2\beta_*$. We thus obtain
\begin{equation}
    S_A = \frac{1}{4G_{3}} \min(D,2\beta_*).
\label{eq:S_A}
\end{equation}
This is plotted in Fig.~\ref{fig:resolution} for a specific value of $\beta_* = \frac{\pi}{3}$. An important point is that the value of $S_A$ given by $\frac{D}{4G_3}$ at $t = t_c$ is $\frac{\pi}{4G_3}$, so that the second QES starts dominating before $t = t_c$ for any choice of region $A$. We also note that if $\beta_* \leq \frac{\pi}{4}$, $S_A$ stays to be a constant value 
$\frac{\beta_*}{2G_3}$ ever since $t = 0$, so to see the transition we need to take region $A$ sufficiently small.

A similar behavior of $S_A$ is obtained in higher dimensions as well. In this case, the saturation value corresponding to $\frac{\beta_*}{2G_3}$ for $d=2$ is given by
\begin{equation}
    S_{A,\infty} = \frac{V_{\bar{A}}}{4G_{d+1}} = \begin{cases}\frac{\pi (1-\cos\beta_*)}{2G_4} & \mbox{for } d=3 \\
    \frac{\bigl(\frac{\pi}{2(d-2)}-\cos{\beta_*}f(d,\beta_*)\bigr)\Sigma_{d-2}}{2G_{d+1}} & \mbox{for } d\geq4
    \end{cases},
\end{equation}
where $V_{\bar{A}}$ is the ($d-1$)-dimensional volume of $\bar{A}$ on $\mbox{dS}_d^2/\mathbb{Z}_{2}$ at $t=0$, $\Sigma_{d-2}$ is the volume of the ($d-2$)-dimensional unit sphare, and
\begin{equation}
  f(d,\beta_*)=\frac{\sqrt{\pi}\Gamma(\frac{d}{2}-1)}{2\Gamma(\frac{d-1}{2})}\, {}_{2}F_{1}\biggl[\frac{1}{2},\frac{3-d}{2},\frac{3}{2},\cos^{2}\!{\beta_*}\biggr].
\end{equation}
The initial value at $t=0$ is given by $S_A = \frac{V_A}{4G_{d+1}}$, where $V_A$ is the volume of $A$ on $\mbox{dS}_d^2/\mathbb{Z}_{2}$.

In the next section, we see that the time dependence of $S_A$ obtained here can indeed be interpreted as a Page curve, reflecting the finite Hilbert space of dS spacetime in quantum gravity.

\section{Interpretation}
\label{sec:interp}

Let us consider what the result obtained in the previous section implies for the fundamental structure of spacetime. We discuss only the nontrivial case of $\beta_* > \frac{\pi}{4}$. We first observe that the initial and final values of $S_A$ can be written as
\begin{alignat}{5}
   S_{A}(t=0) &= \frac{V_{A}}{4G_{d+1}} &&= \frac{\mathcal{A}}{4G_d} \biggl(\frac{V_A}{V}\biggr),
\label{eq:SA-init}\\
    S_{A}(t \rightarrow \infty) &= \frac{V_{\bar{A}}}{4G_{d+1}} &&= \frac{\mathcal{A}}{4G_d} \biggl(\frac{V_{\bar{A}}}{V}\biggr),
\label{eq:SA-fin}
\end{alignat}
where
\begin{equation}
    G_d \equiv \frac{\cal A}{V} G_{d+1},
\end{equation}
and $V$ and ${\cal A}$ are the volume and horizon area of $\mbox{dS}_d/\mathbb{Z}_{2}$ at $t = 0$ (i.e., the area of the hemisphere and the length of the half equator in the right panel of Fig.~\ref{fig:Z2}), respectively. It is easy to see that $G_d$ is, indeed, Newton's constant in the $d$-dimensional theory on $\mbox{dS}_d^2/\mathbb{Z}_{2}$.

To decipher the meaning of Eqs.~(\ref{eq:SA-init},~\ref{eq:SA-fin}), we take the ``original'' boundary description of double holography discussed at the end of Section~\ref{sec:embedding}. In this description, the total system is a gravitational system on ${\cal M} \equiv \mbox{dS}_d^2/\mathbb{Z}_{2}$ with boundary $\partial {\cal M}$. In the ($d+1$)-dimensional bulk picture, $\partial {\cal M}$ is the intersection of ${\cal M}$ and $Q$, and the number of degrees of freedom associated with it is related to the tension of the $Q$ brane~\cite{Takayanagi:2011zk,Fujita:2011fp}. In our case, the $Q$ brane is tensionless, implying that $\partial {\cal M}$ does not carry its own degrees of freedom at the leading order.

This allows us to interpret Eqs.~(\ref{eq:SA-init},~\ref{eq:SA-fin}) that the initial and final values of $S_A$ are given, respectively, by the fractions of the spatial volume which $A$ and $\bar{A}$ occupy at $t=0$, multiplied by the quantity $\frac{\cal A}{4G_d}$, the Gibbons-Hawking entropy of the spacetime $\mbox{dS}_d/\mathbb{Z}_{2}$. Is this interpretation reasonable?

The interpretation just described, in fact, is consonant with the picture of spacetime developed in Refs.~\cite{Nomura:2018kia,Nomura:2019qps,Nomura:2019dlz,Nomura:2020ska,Nomura:2020ewg}. Let us consider a spacetime with a black hole. A ``fundamental,'' unitary description of this system is obtained by viewing the black hole from a distance~\cite{tHooft:1990fkf,Susskind:1993if}, in which the Bekenstein-Hawking entropy is interpreted as the logarithm of the number of independent microstates associated with the black hole spacetime. The relevant degrees of freedom, called soft modes~\cite{Nomura:2018kia,Nomura:2019qps}, are localized mostly on the stretched horizon~\cite{Susskind:1993if}. The picture of the black hole interior (or a two-sided black hole, corresponding to the maximally extended spacetime in general relativity) arises as a collective phenomenon, with the resulting vacuum state being the unique semiclassical vacuum state because of the coarse graining employed in obtaining the picture~\cite{Nomura:2019dlz,Nomura:2020ska,Nomura:2020ewg}.

A similar construction also applies to dS spacetime~\cite{Nomura:2019qps}, though it is more speculative. In this case, the description based on the static coordinates corresponds to the distant picture of a black hole. The Gibbons-Hawking entropy represents the logarithm of the number of microstates of the dS spacetime, and the degrees of freedom representing them, i.e.,\ soft modes, are located mostly on the (stretched) horizon. This localization of soft modes occurs because of a large gravitational blueshift associated with the existence of the horizon.

The extension of the dS spacetime covering the full spatial section at $t=0$ can be performed through coarse graining~\cite{Nomura:2019qps}, analogous to the black hole case. In the extended space, the vacuum state is the unique semiclassical dS vacuum state, regardless of the microstate we started from. (In fact, the state has a very special, antipodal entanglement structure at $t=0$~\cite{Miyaji:2015yva,Geng:2019ruz,Geng:2020kxh}.) The concept of vacuum microstates does not exist in the (maximally) extended theory because of the coarse graining involved in erecting such a theory~\cite{Nomura:2020ewg,Nomura:2018kia,Nomura:2019qps}.

Nevertheless, there is a concept of microscopic vacuum {\it degrees of freedom}, even though they are in the unique vacuum {\it state}. How are these degrees of freedom distributed in space? Since the extended geometry at $t=0$ is a homogeneous space with no special point selected, it is natural to expect that the number of degrees of freedom indicated by the Gibbons-Hawking entropy is distributed uniformly throughout the space. In this picture, the Gibbons-Hawking entropy represents the entanglement entropy between two subregions when the dS space at $t=0$ is divided into two hemispheres. The whole discussion given here also goes through in our setup with the $\mathbb{Z}_{2}$ identification if the appropriate adjustments are made; for example, the entanglement surface in the last sentence must be taken to be orthogonal to the fixed surface of $\mathbb{Z}_{2}$ (as depicted in Fig.~\ref{fig:Z2}), and the value of the entropy must be divided by a factor of~$2$.

The picture described above is precisely what is suggested by our result. Given that dS spacetime at $t=0$ is a homogeneous, scrambled system, the entropy of subregion $A$ at $t=0$ is given by the number of degrees of freedom there:
\begin{equation}
    n_{A,0} = \frac{S_{\rm GH}}{2} \biggl(\frac{V_A}{V}\biggr) = \frac{\cal A}{4G_d} \biggl(\frac{V_A}{V}\biggr),
\label{eq:Hilbert-A}
\end{equation}
where $S_{\rm GH}$ is the Gibbons-Hawking entropy, and the factor $2$ in the denominator of the middle expression comes from the $\mathbb{Z}_{2}$ identification. (Note that we are taking subregion $A$ smaller than $\bar{A}$ at $t=0$.) We emphasize that the spatial distribution of the microscopic degrees of freedom appears very different between the static and global pictures (between the boundary and two-sided pictures in the case of a collapse-formed black hole).

The initial increase of $S_A$ implies that the number of degrees of freedom in $A$ increases in time. This is possible because the number of degrees of freedom captured by the holographic entropy is, in fact, that of {\it effective} degrees of freedom necessary to describe the emergent bulk spacetime, which is determined by the entanglement structure of (presumably an infinite number of) more fundamental degrees of freedom and can thus vary in time~\cite{Miyaji:2015yva,Nomura:2018kji,Murdia:2020iac}.
In our context, the increase of $S_A$ arises from the emergence of the ``interior'' region, i.e.,\ the upper quadrant of the Penrose diagram in Fig.~\ref{fig:ext-st}, analogous to the Hartman-Maldacena setup~\cite{Hartman:2013qma}.

This increase of $S_A$, however, does not last forever, since subregion $\bar{A}$ does not ``sense'' the emergence of the interior, and hence the number of degrees of freedom there,
\begin{equation}
    n_{\bar{A}} = \frac{S_{\rm GH}}{2} \biggl(\frac{V_{\bar{A}}}{V}\biggr) = \frac{\cal A}{4G_d} \biggl(\frac{V_{\bar{A}}}{V}\biggr),
\label{eq:Hilbert-Abar}
\end{equation}
stays constant. If the state of the total system is pure, as we assume here, then the entanglement entropy of $A$ cannot become larger than this number. This is the origin of the saturation of $S_A$ we have seen, which is nothing but the Page transition phenomenon representing unitarity of the underlying dynamics.

\section{Conclusions}
\label{sec:concl}

In this paper, we have formulated a version of the information paradox in dS spacetime and shown that it is solved by the emergence of entanglement islands in the context of double DS/dS holography. In particular, we have obtained a time-dependent Page curve in a setup in which the graviton stays massless.

Beyond the specific context of dS spacetime, our result suggests the following:
\begin{itemize}
\item The entanglement entropy of Hawking radiation appearing in the calculation of the Page curve can be computed by the QES prescription by fixing a region in which the radiation resides, even if it is in a (weakly) gravitating region.
\item While the Gibbons-Hawking/Bekenstein-Hawking entropy represents the logarithm of the number of independent dS/black-hole microstates in a static/distant description (i.e., in unitary gauge construction~\cite{Nomura:2020ewg,Langhoff:2020jqa}), the vacuum state in an analytically extended spacetime---a global/two-sided description---is a unique, semiclassical state.
\item In the static/distant description of a dS/black-hole spacetime, the degrees of freedom associated with the microstates are mostly localized near the horizon, due to a strong gravitational blueshift. On the other hand, in the analytically extended spacetime, they are distributed uniformly in space (at the time the extension is made).
\item The increase of volume in the ``interior'' region of dS spacetime (the top quadrant of the Penrose diagram) has a physical meaning in the sense that the entanglement entropy of a region surrounding it increases in time (until it is saturated by the Page transition).
\end{itemize}
We hope that these findings shed further light on how spacetime and gravity work at the fundamental level.

\begin{acknowledgments}
We thank Andreas Karch, Suvrat Raju, Lisa Randall, and Pratik Rath for useful discussions. The work of Y.N. was supported in part by the Department of Energy, Office of Science, Office of High Energy Physics under contract DE-AC02-05CH11231 and award DE-SC0019380 and in part by MEXT KAKENHI grant number JP20H05850, JP20H05860. The work of H.-Y.S. was supported by the Simons Collaborations on Ultra-Quantum Matter, grant 651440 (AK) from the Simons Foundation. H.G. is very grateful to his parents and recommenders.
\end{acknowledgments}

\bibliographystyle{apsrev4-1}
\bibliography{DSEG}
\end{document}